\documentclass{epl}
\usepackage{epsfig}

\begin{document}  

\newcommand{\prl}{Phys.\ Rev.\ Lett.}
\newcommand{\pra}{Phys.\ Rev. A}
 
\title{Collisions and expansion of an ultracold dilute Fermi gas}

\author{B. Jackson\inst{1}, P. Pedri\inst{1,2}, and 
S. Stringari\inst{1}}
\shortauthor{B. Jackson \etal} 
\institute{
 \inst{1} Dipartimento di Fisica, Universit\`a di Trento and BEC-INFM, 
 I-38050 Povo, Italy \\
 \inst{2} Institut f\"ur Theoretische Physik, Universit\"at Hannover, 
 D-30167 Hannover,Germany}


\maketitle  

\begin{abstract}  

We discuss the effects of collisions on the expansion of a degenerate normal
Fermi gas, following the sudden removal of the confining trap. Using a 
Boltzmann equation approach, we calculate the time dependence of the aspect 
ratio and the entropy increase of the expanding atomic cloud taking into 
account the collisional effects due to the deformation of the distribution 
function in momentum space. We find that in dilute gases the 
aspect ratio does not deviate significantly from the predictions of ballistic 
expansion. Conversely, if the trap is sufficiently elongated the thermal 
broadening of the density distribution due to the entropy increase can be 
sizeable, revealing that even at zero temperature collisions are effective in 
a Fermi gas.

\end{abstract}  

The expansion of an atomic gas following the sudden removal of the
confining trap is known to provide valuable information on the state of
the system and on the role of interactions \cite{rmp}. In particular
in the superfluid
phase, where the macroscopic dynamics of the gas are governed by the
equations of hydrodynamics, one expects anisotropic expansion if the 
gas is released from an anisotropic trap. This peculiar feature, which should
be contrasted with the isotropic ballistic expansion exhibited by
a non interacting gas, was first discussed in
\cite{castin,kagan} in the case of Bose gases and in \cite{menotti} in the
case of Fermi gases. Anisotropy is not however a unique feature exhibited
by superfluids and also in the normal phase one can expect a similar
behaviour if collisions are sufficiently important 
\cite{kagan,arimondo,svarchuck,gerbier}. This
effect is expected to be particularly important in the case of 
Fermi gases interacting with large scattering
lengths near a Feshbach resonance \cite{thomas,jin,salomon}. At first sight one
would expect that the effects of collisions are suppressed at low
temperature because of Pauli blocking. This is certainly true if one
works close to equilibrium,
as happens in the study of small amplitude oscillations \cite{Vichi}. In
the problem of the expansion, however, large deformations in momentum
space can be produced if one starts from a highly deformed cloud, with
the result that collisions become effective
even if the gas is initially at zero temperature. This
interesting possibility was
first pointed out in \cite{Anglin}.

The aim of the present work is to calculate explicitly the dynamics of the
expansion of a dilute, degenerate normal Fermi gas, taking into account the
role of collisions. The main purpose is to provide quantitative predictions
for the aspect ratio and the thermal broadening of the density distribution, 
as a function of the relevant parameters like the
ratio of the trap
frequencies, the scattering length and the number of atoms.

The system we consider is a dilute two component Fermi gas. At low
temperatures the collisions between two atoms of the same species are 
suppressed
due to Pauli blocking, and only atoms of different species can collide. We 
assume that the two species have the same mass and density. The starting 
point is the Boltzmann equation
\begin{equation}
\label{eq:Bol}
 \frac{\partial f}{\partial t}+{\bf v}_1\cdot\nabla_{\bf r}f
 -\frac{1}{m} \nabla U_{\rm ext} \cdot\nabla_{{\bf v}_1}f=C[f],
\end{equation} 
where for each component $f({\bf r},{\bf v}_1,t)$ is the distribution function 
in phase-space and $U_{\rm ext}({\bf r})=(m/2)(\omega_\perp^2 (x^2+y^2)+
\omega_z^2 z^2)$ is the external trap potential. We note that the potential
is cylindrically symmetric, which is the form favoured by experiments. 
Moreover, in (\ref{eq:Bol}) we have neglected the mean field interaction term 
\cite{David} which, however, has a minor effect on the expansion of a dilute 
Fermi gas \cite{menotti}. The collisional integral in the case of a dilute Fermi
system reads
\begin{eqnarray}
C[f]=\frac{\sigma m^3}{4\pi h^3}\int d^3v_2 \, d^2 \Omega\, |{\bf v}_1-{\bf
v}_2|[(1-f({\bf v}_1))(1-f({\bf v}_2))f({\bf v}'_1)f({\bf v}'_2) \nonumber \\
-f({\bf v}_1)f({\bf v}_2)(1-f({\bf v}'_1))(1-f({\bf v}'_2))],
\end{eqnarray}  
where in the low-energy limit, $\sigma=4\pi a^2$, with $a$ the s-wave
scattering length. The scattering length is assumed to be smaller than the
average distance between atoms. 

The dynamics can be studied analytically using the
method of the averages \cite{David2,cla_exp} which involves calculating moments 
of the type
$\langle\chi\rangle(t)=\int\chi({\bf r},{\bf v})f({\bf r},{\bf v},t)
 \, d^3r \,d^3v/\int f({\bf r},{\bf v},t)\, d^3r\, d^3v$.
Applying this to 
(\ref{eq:Bol}) gives a set of equations of the general form
\begin{equation}
 \frac{d \langle\chi\rangle}{dt}-\langle{\bf v}\cdot\nabla_{\bf r}\chi\rangle
 - \frac{1}{m} \langle \nabla U_{\rm ext} \cdot\nabla_{\bf v}\chi\rangle=
 \langle \chi C[f]/f\rangle,
\end{equation}
where $\chi$ is a general
function which depends on ${\bf r}$ and ${\bf v}$. Here we shall take
moments of the form $\langle r_i^2 \rangle$ , $\langle r_i v_i \rangle$ 
and $\langle (v_i-u_i)^2 \rangle$, where $u_i({\bf r})$ represents the
velocity fields within the gas. 
If $\chi$ is a conserved quantity during a collision (i.e.\ 
$\Delta \chi=\chi_1+\chi_2-\chi_{1'}-\chi_{2'}$ is zero) then
the collisional term disappears. This is 
true for the moments $\langle r_i^2 \rangle$ and $\langle r_i v_i \rangle$
as well as for the quantity 
$\sum_i \langle (v_i-u_i)^2 \rangle$. This method has been used to 
study collective excitations in \cite{Vichi}.

To explicitly include the collisional term in the calculations, 
it is convenient to introduce the following parametrization for the 
distribution function
\begin{equation}
\label{eq:f-ansatz}
 f({\bf r},{\bf v},t)=
 \left[\exp\left(\frac{E({\bf r},{\bf v},t)-\mu}{T}\right)+1\right]^{-1},
\end{equation}
with
\begin{equation}
 E({\bf r},{\bf v},t)=\frac{m}{2} \sum_i \left[ \frac{\omega_i^2 r_i^2}{b_i^2}
 + \frac{(v_i-u_i)^2}{K_i} \right],
\label{eq:scaling}
\end{equation}
where $u_i=\beta_i r_i$, and $\beta_i$, $b_i$, $K_i$, $\mu$ and $T$ are time 
dependent parameters, with $i\in\{x,y,z\}$. The ansatz
(\ref{eq:f-ansatz},\ref{eq:scaling}) contains more free parameters than 
necessary, so that we can, without any loss of generality, set
$\prod_i b_i=(\prod_i K_i)^{-1/2}$. This fixes a unique definition of the 
effective temperature $T$ \cite{note0}. The chemical potential, $\mu$, is then
obtained from the normalization condition, $(T/\hbar \bar{\omega})^3 f_3
 ({\rm e}^{\mu/T}) = N/2$,
where $N$ is the total number of atoms in {\em both} components, 
$\bar{\omega}^3=\prod_i \omega_i$, and $f_s(z)=\sum_n (-1)^{n+1} z^n/n^s$. 
From Eqs.(\ref{eq:f-ansatz},\ref{eq:scaling}) one can evaluate the entropy 
per particle of the gas, $S=(4\rho-\mu)/T$, where
$\rho=T f_4 ({\rm e}^{\mu/T})/f_3 ({\rm e}^{\mu/T})$. 
This result, taken together with the normalization condition, shows that 
there is a one-to-one correspondence between the entropy and the temperature 
$T$, and consequently changes in $T$ taking 
place during the expansion are associated with a change of entropy.

The ansatz (\ref{eq:f-ansatz},\ref{eq:scaling}) includes the initial 
equilibrium configuration predicted
by Fermi statistics, and accounts for rescaling effects in coordinate as
well as in momentum space. It allows, in particular, for anisotropic
effects in momentum space which are crucial for describing correctly the
mechanism of the expansion. The parametrization can describe
both the collisionless and the hydrodynamic expansion as well as other
intermediate regimes. Furthermore it accounts for possible changes in the
effective temperature of the system during the expansion.

The next step is to evaluate the moments $\langle \chi \rangle$ in terms of the
scaling parameters, which gives
$\langle r_i^2 \rangle =b_i^2 \rho / m \omega_i^2$. Other moments can be 
expressed
in terms of this quantity, so that $\langle r_i v_i \rangle = \beta_i 
\langle r_i^2 \rangle$, $\langle (v_i -u_i)^2 \rangle =\omega_i^2 K_i   
\langle r_i^2 \rangle/b_i^2$, and $\langle u_i^2 \rangle = \beta_i^2 
\langle r_i^2 \rangle$. The moment equations then become 
\begin{eqnarray}
\label{eq:continuity}
&&\beta_i=\frac{\dot{b}_i}{b_i}+\frac{\dot{\rho}}{2\rho}, \\
\label{eq:2}
&&\dot{\beta}_i + \beta_i^2 - \frac{\omega_i^2}{b_i^2} K_i +\omega_i^2= 0, \\
\label{eq:3}
&&\frac{\dot{K}_i}{K_i}-2\frac{\dot{b}_i}{b_i}+4\beta_i =  
 \frac{\langle (v_i-u_i)^2 C/f \rangle}{\langle (v_i-u_i)^2 \rangle},
\end{eqnarray}
where the final term in (\ref{eq:2}) arises from the confining potential. We
set this term to zero in studying the expansion of the gas, while rescaling 
time in units of $\omega_{\perp}$, the original radial trap frequency. 
In addition, it is convenient to rewrite 
(\ref{eq:3}) as equations for the anisotropy in momentum space, 
$s=(K_z/K_\perp)^{1/2}$, and the entropy, $S$
\begin{eqnarray}
\label{eq:fin-5}
&&\frac{\dot{s}}{s}=\frac{\dot{b}_{\perp}}{b_{\perp}}-\frac{\dot{b}_z}{b_z}
 -\frac{(2+s^2)}{4s^2} K^2 \xi J (s,\tau), \\
\label{eq:fin-6}
&&\frac{T}{\rho} \dot{S}+ \frac{(1-s^2)}{2s^2} K^2 \xi J (s,\tau)=0.
\end{eqnarray} 
where we have defined
\begin{equation}
 \frac{1}{\omega_{\perp}} \frac{\langle (v_{\perp}-u_{\perp})^2 C/f\rangle}
 {\langle (v_{\perp}-u_{\perp})^2 \rangle}= 
 K^2 \xi J \left(s,\tau \right),
\label{eq:collterm}
\end{equation}
with $J(s,\tau)$ given by
\begin{equation}
 J (s,\tau) = \frac{3^{1/3}}{4\pi^6} \tau^2
 \frac{s^{5/3}}{f_4 ({\rm e}^{\mu/T})} \int d^3 R \, d^3 V_1 \, 
 d^3 V_2 \, d^2 \Omega \, V' \Delta V_z^2
 f({\bf V}_1)f({\bf V}_2)(1-f({\bf V}'_1))
 (1-f({\bf V}'_2)).
\label{eq:jrt}
\end{equation}
Here $V'=\sqrt{(V_{1x}-V_{2x})^2+(V_{1y}-V_{2y})^2+s^2(V_{1z}-V_{2z})^2}$, 
and we have rescaled variables so that 
$f({\bf V})=[\exp (R^2+V^2-\mu/T)+1]^{-1}$. 
The function $J$ depends on both $s$ and 
the reduced temperature $\tau=T/T_F$, where $T_F=(3N)^{1/3} \hbar 
\bar{\omega}$ is the Fermi temperature.
In Eqs.\ (\ref{eq:fin-5}-\ref{eq:collterm}) we have introduced the relevant 
dimensionless interaction parameter
\begin{equation}
\xi=(\lambda N)^{1/3} (k_F a)^2,
\end{equation}
where $k_F$ is the initial Fermi
wavevector at the center of the trap, and $\lambda=\omega_z/\omega_{\perp}$
is the trap anisotropy. Further, we have introduced the geometric average 
$K=(K_\perp^2 K_z)^{1/3}=(b_\perp^2 b_z)^{-2/3}$. 

Eq.\ (\ref{eq:fin-6}) explicitly shows that the entropy (and hence the 
temperature) of the gas will 
remain constant during the expansion ($\dot{S}=0$) either in the absence of 
collisions ($J=0$) or when the distribution in momentum space is isotropic 
($s=1$) \cite{note0b}. The latter situation arises when starting from a 
spherical trap, 
$\lambda=1$, or when collisions are sufficiently frequent so that the gas is in
the hydrodynamic regime. In regimes intermediate between the collisionless and
hydroynamic limits, deformations in momentum space produce an increase of 
entropy and temperature. 
The equations 
(\ref{eq:continuity}-\ref{eq:jrt}) can also be used to 
study small oscillations around equilibrium, where the momentum distribution 
is isotropic ($s=1$). This procedure reproduces exactly the 
results of \cite{Vichi}, in particular the collisional term takes the 
form $\langle (v_i-u_i)^2 C/f\rangle/\langle (v_i-u_i)^2 \rangle = 
(s-1)/\tilde{\tau}$, with $1/(\omega_\perp \tilde{\tau})=\mathcal {C} 
K^2 \xi F_Q (T/T_F)$ and $\mathcal{C}=8/(5\cdot 3^{5/3})$. Here 
$\tilde{\tau}$ is the typical relaxation time 
for the shape oscillations and $F_Q$ is given in \cite{Vichi}.
Note that $1/\tilde{\tau} \propto T^2$ at small temperatures, so that the 
collisional contribution disappears when $T=0$. This is the result of Pauli 
blocking of collisions for a spherical momentum
distribution. However, during expansion large deformations in
momentum space can lead to collisions that scatter atoms outside of the Fermi 
surface, and the function $J$ differs from zero, even at zero temperature, as
pointed out in \cite{Anglin}.

\begin{figure}[here]
\centering
 \scalebox{.47}
 {\includegraphics{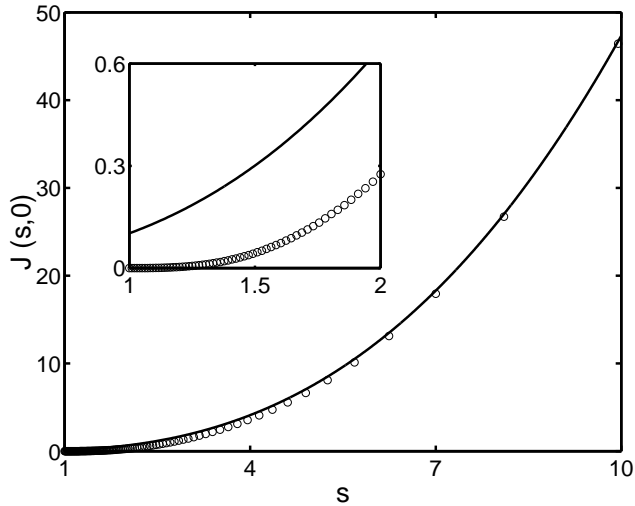}}
\caption{$J(s,0)$ as a function of $s$ for a range between 
 $1$ and $10$, and (inset) between $1$ and $2$. The solid line shows the 
 analytical
 form $J(s) = 2s^{8/3}/(3^{5/3} \pi)$ for the limit $s \rightarrow \infty$.}
\label{fig:jr}
\end{figure}

We will assume that the gas is initially in equilibrium at
low temperature, and that
the effective temperature contained in the ansatz (\ref{eq:f-ansatz}) remains 
small during the expansion. One can then expand the Fermi functions $f_s(z)$ 
using low-$T$ Sommerfeld expansions, so that the normalization condition
gives $\mu \simeq T_F (1-\pi^2 \tau^2/3)$,
$\rho \simeq T_F( 1+2\pi^2\tau^2/3)/4$ and $S=\pi^2 \tau$.
Substituting these expressions into
(\ref{eq:continuity},\ref{eq:2},\ref{eq:fin-5},\ref{eq:fin-6}) yields a new set
of equations that can be solved numerically to study the time evolution of the 
gas. To simplify matters we also approximate the function $J(s,\tau)$ 
with its zero temperature value
\begin{equation} 
 J (s,0) = \frac{2 \cdot 3^{4/3}s^{5/3}}{\pi^6}\int d^3 R \, d^3 V_1 \, 
 d^3 V_2 \, d^2 \Omega \, V' \Delta V_z^2 
 \theta_0({\bf V}_1)\theta_0({\bf V}_2)(1-\theta_0({\bf V}'_1))
 (1-\theta_0({\bf V}'_2)),
\label{eq:zerotemp}
\end{equation} 
where we have rescaled the coordinates again
such that $\theta_0 ({\bf V})=\Theta(1-V^2-R^2)$, where $\Theta$ is the
Heaviside step function.
The integral for $J(s)$ can be evaluated numerically using a standard Monte
Carlo technique, and is plotted in 
Fig.\ \ref{fig:jr}. Also plotted is the
result $J(s) = 2 s^{8/3}/(3^{5/3} \pi)$ holding for large $s$, where 
the integral can be 
evaluated analytically by noting that 
virtually all collisions in this limit will scatter atoms outside of the Fermi 
surface (i.e.\ $(1-\theta_0({\bf V}'_1))(1-\theta_0({\bf V}'_2)) \simeq 1)$. 
We shall discuss the validity of this zero-temperature approximation later.

\begin{figure}[here]
\centering
 \scalebox{.72}
 {\includegraphics{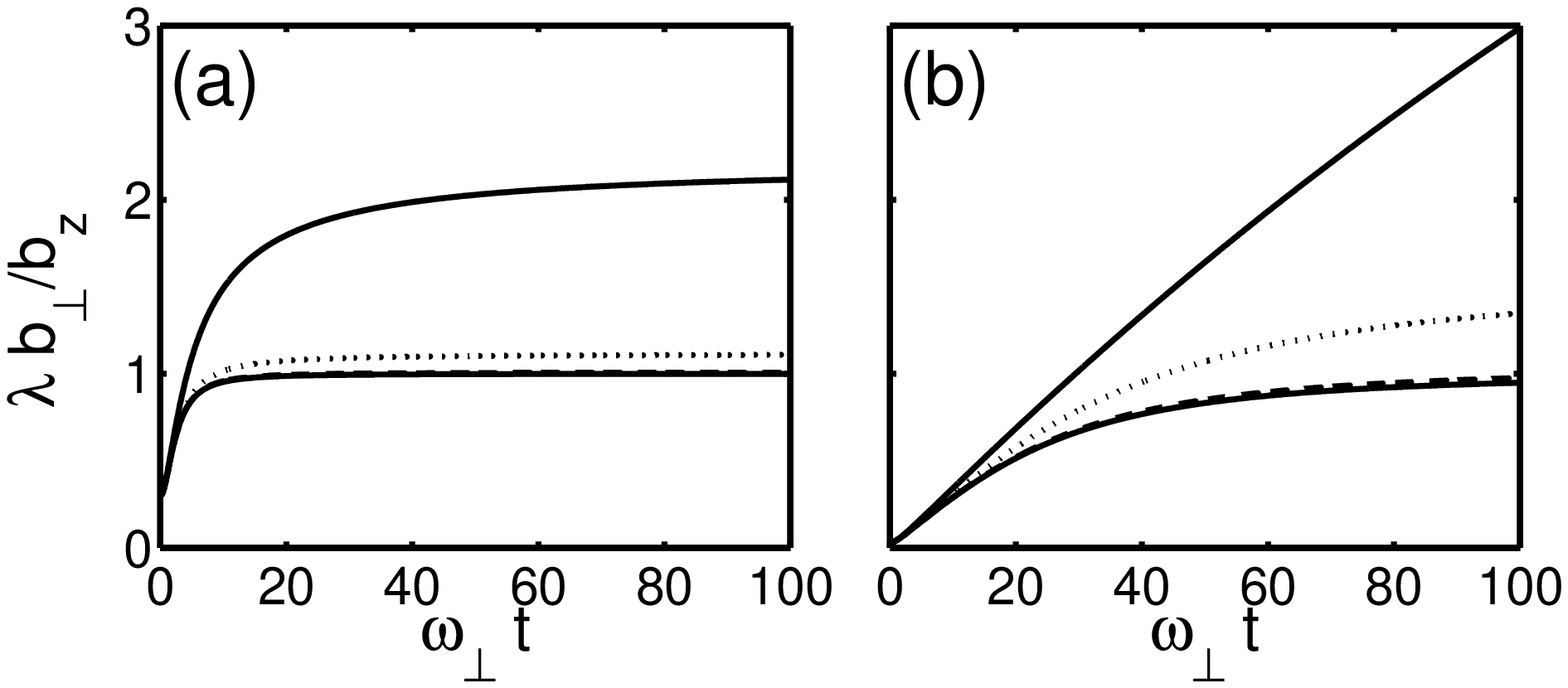}}
\caption{Aspect ratio against time for the free expansion of a 
 zero-temperature Fermi gas from a trap with 
 (a) $\lambda=0.3$ and (b) $\lambda=0.03$. The upper and lower solid lines on
 each plot represents the behavior in the hydrodynamic and collisionless 
 regimes respectively, with the other lines displaying results of solving
 the equations (\ref{eq:continuity}-\ref{eq:jrt}) for $\xi=1$ (dashed) and 
 $\xi=20$ (dotted).}
\label{fig:aspecr}
\end{figure}

We now use these equations to study the expansion for different choices of the 
trap anisotropies, $\lambda<1$, starting from a $T=0$ 
configuration \cite{note1}. We plot the aspect ratio 
of the cloud $\sqrt{\langle r_{\perp}^2 \rangle/\langle z^2 \rangle}=
\lambda b_\perp/b_z$ as a function of time for 
$\xi=1$ and $\xi=20$ and compare to the expected 
collisionless and hydrodynamic behavior in Fig.\ \ref{fig:aspecr}. One notices 
that the effects of collisions are far less pronounced for $\lambda=0.3$, 
where the results are barely distinguishable from 
the collisionless behavior. However, for $\lambda=0.03$ the expansion
lies intermediate between the two limits if one chooses $\xi=20$. This is to be
expected since the deformation in momentum space reached in this case is much
larger. In particular, in the collisionless case one would expect 
$s \rightarrow 1/\lambda$ at long times, and this fixes the maximum $s$ 
possible. Since $J$ takes large values only for large $s$, then the collisional
term is potentially more effective for smaller $\lambda$. 

\begin{figure}[here]
\centering
 \scalebox{.66}
 {\includegraphics{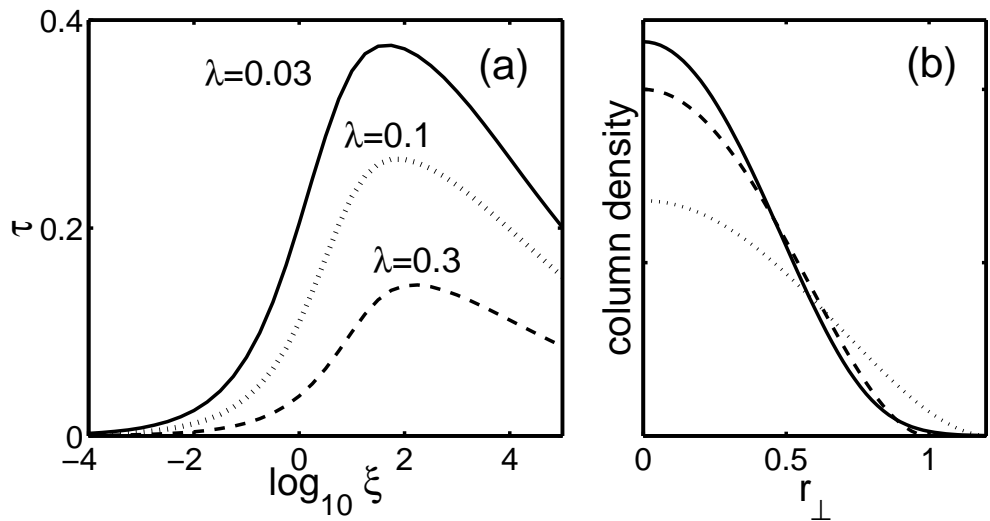}}
\caption{(a) Temperature at $\omega_\perp t=100$, $\tau=T/T_F$, 
 as a function of $\xi$, for different values of $\lambda$.
 The gas prior to expansion is at zero temperature.(b) Column density 
 (see text) as a function of radial distance at
 $\omega_{\perp} t=100$, for $\lambda=0.03$ and $\xi=1$ (solid line), compared 
 to the result of expansion in the collisionless (dashed) and 
 hydrodynamic (dotted) limits. Both axes are in arbitrary units.}
\label{fig:temper}
\end{figure}

An interesting consequence of the momentum space deformation is that 
atoms scattering outside the Fermi surface will tend to smooth out the 
distribution function. This aspect appears in 
(\ref{eq:fin-6}) as an increase in the temperature, even if we start from an
initial $T=0$ configuration. Fig.\ \ref{fig:temper}(a) 
shows the temperature calculated at $\omega_{\perp} t=100$ for different 
values of $\lambda$. Again, one sees a much larger effect for 
$\lambda\ll 1$. 
The function tends to zero for both $\xi \rightarrow 0$ and 
$\xi \rightarrow \infty$, representing the crossover from collisionless 
to hydrodynamic behavior in a similar manner to damping times in collective
oscillations \cite{Vichi,David2}. Since (\ref{eq:fin-6}) is an equation for
$\tau^2$ one finds that starting from a non-zero 
temperature, $\tau_0$, gives a final temperature of approximately 
$\tau=\sqrt{\tau_0^2+(\delta \tau)^2}$, where $\delta \tau$ is the change 
for an initially zero temperature gas obtained from Fig.\ \ref{fig:temper}(a).
Hence the temperature increase produced by collisions will become
smaller as one raises the initial temperature of the gas.  

The change in temperature during the expansion, as well as the fact that 
experiments will be 
initially at a non-zero temperature, will lead to corrections to the 
zero-temperature collisional term (\ref{eq:zerotemp}). We can estimate the
effects at finite temperatures by approximating $J(s,\tau)$ (\ref{eq:jrt}) 
with the sum of the zero-$T$ result to that derived for finite-$T$ but small 
deformations, so that $J(s,\tau)=J(s,0)+\mathcal{C} (s-1)F_Q(\tau)$, with the 
function $F_Q (\tau)$ given in \cite{Vichi}. We find that the inclusion of the
temperature dependence in $J$ has little impact on the results, even allowing
for an initial temperature of $\tau=0.2$. 

Since for a dilute gases ($k_F |a| \ll 1$) the value of $\xi$ for realistic
parameters will be, at most, of the order of 1, we conclude that collisional 
effects on the aspect ratio of the expanding gas are negligible (see Fig.\
\ref{fig:aspecr}). This is 
consistent with experimental results in dilute degenerate gases (see for 
example \cite{jin,demarco}). In contrast, we find that
the effect of collisions on the thermal broadening can be significant at low 
temperatures even if $\xi \simeq1$, especially for highly deformed traps. 
As an example in Fig.\ \ref{fig:temper}(b) we show the column density 
$n(r_{\perp}) = (m/h)^3 \int dz d{\bf  v} f({\bf r},{\bf v},t)$, 
calculated at $\omega_{\perp}t=100$, $\lambda=0.03$ and $\xi=1$, starting from 
an initial zero temperature configuration. Around $80\%$ of the increase 
in $T$ (and the consequent broadening of the density profile) takes place
over the first $\omega_{\perp} t =20$ of the expansion. The comparison with 
the prediction 
of ballistic expansion (dashed line) explicitly reveals the importance of 
collisions. Experimentally one could observe this difference by cooling down
a two component Fermi gas to very low temperatures. Ballistic expansion could 
be achieved either by first removing one of the two components, or by
suddenly tuning the scattering length to zero at the start of the expansion.

In order to also observe large effects in the aspect ratio one should increase
the value of $\xi$, and hence of $k_F |a|$, by, for example working close to a 
Feshbach resonance. In this case, however, the
formalism of the Bolzmann equation is not strictly applicable.
A rough estimate of the collisional effects can be obtained by
replacing $a^2$ with the unitarity limited expression $a^2/(1+K(k_F a)^2)$, 
where $k_F$ is the initial Fermi momentum and $K$ accounts for the decrease 
of the density during the expansion. In the unitarity limit 
$k_F |a| \rightarrow \infty$ equation (\ref{eq:collterm}) is then modified by
replacing $K^2$ with $K$ and setting $\xi=(\lambda N)^{1/3}$. These changes 
result in a sizable increase of the anisotropy effects. Apart from the fact
that the parameter $\xi$ can easily take large values, the 
replacement of $K^2$ with $K$ makes the collisional term effective for longer 
times during the expansion. One should however note that in the unitarity 
limit the gas is expected to be superfluid at low temperatures and its dynamics
should be consequently described by the hydrodynamic equations of superfluids.
  
In conclusion we have shown that collisions can be effective in a dilute 
normal Fermi gas even at zero temperature, as a consequence of large 
deformations of the distribution function in momentum space after 
expansion from a very elongated trap. They can give rise to a sizeable 
entropy increase and hence thermal 
broadening of the density distribution, which should be visible by imaging the 
atomic cloud. In contrast a $T=0$ superfluid should expand anisotropically,
without any entropy increase due to the absence of collisions. 

\acknowledgments

We acknowledge support from Deutsche Forschungsgemeinschaft (SFB 407),  
the RTN Cold Quantum gases, ESF PESC BEC2000+, and the Ministero 
dell'Istruzione, dell'Universit\`a e della Ricerca (MIUR).
P. P. wish to thank the Alexander von Humboldt Foundation, 
the Federal Ministry of Education and 
Research and the ZIP Programme of the German Government.


\begin{thebibliography}{0}
\bibitem{note0} Alternatively, one can fix the value of $T$ to the initial
 value and obtain an equivalent set of equations, as in the work of Pedri 
 \etal \cite{cla_exp}. The choice made in the present work is particularly
 convenient if the expansion starts from a $T=0$ configuration.
\bibitem{note0b} Note that Eq.\ (\ref{eq:fin-6}) can also be obtained by 
 starting strictly from the Boltzmann H-theorem.
\bibitem{note1} In practice, the $T=0$ approximation is compatible with the 
 assumption that the gas is non superfluid, the critical temperature for the 
 BCS transition being exponentially small in a dilute gas.
\end{thebibliography}
\end{document}